# Laser-induced Phase-change Perovskite Photodetector Arrays for Optical Information Storage and Imaging

*Chen Zou[1], Jiajiu Zheng[1], Cheng Chang[1], Arka Majumdar[1,2], Lih Y. Lin[1]\**

[1]Department of Electrical and Computer Engineering, University of Washington, Seattle, WA 98195, USA

[2]Department of Physics, University of Washington, Seattle, WA 98195, USA

*Corresponding author. E-mail: lylin@uw.edu

**Abstract**

The reversible phase transition between orthorhombic and cubic phases is demonstrated in all-inorganic perovskite $CsPbIBr_2$ grown by a dual-source vapor-deposition method. The phase transition is actuated via heating and moisture exposure. The different crystal structures of two phases result in distinct optoelectronic properties including optical absorption, refractive index, and carrier transport. The perovskite photodetector array for the non-volatile rewritable optical memory application is further investigated. A near-infrared (NIR) laser ($\lambda = 1064$ nm) is used to selectively heat photodetector pixels through the photothermal effect of the interfacial Ag electrodes, resulting in an optically absorbing perovskite phase for the photodetectors. The locations of specific laser-written pixels can be read out by measuring their photocurrents, and the stored optical information can also be erased by moisture exposure. Finally, a proof-of-concept optical imaging application has been demonstrated using perovskite photodetector arrays fabricated on flexible PET substrates. These results show promising applications of vapor-deposited inorganic perovskite for optical memory and image sensors, with unique potential for low-cost manufacturing of large-area and/or flexible devices.

**Introduction**

Metal halide perovskite has attracted much research interest in recent years due to its outstanding optoelectronic properties including excellent optical absorption, high carrier mobility and long carrier lifetime.[1-2] Since the first perovskite solar cell was reported,[3] significant research efforts have been devoted to improve the photovoltaic efficiency and understand the optical and electronic properties of the underlying material, viz. metal halide perovskite.[4-5] In the last few years, the power conversion efficiency (PCE) of organic-inorganic perovskite solar cells has rapidly advanced to >22%.[6] The crystal structure of metal halide perovskites has been a topic of interest to explain their unique optoelectronic properties.[7-8] The methylammonium metal halide perovskites (MAPbX$_3$), in which the organic cation (MA$^+$) fills up the voids of corner-sharing PbI$_6$ octahedra framework, were first intensively investigated,[9] and MAPbI$_3$ was reported to undergo phase transition from tetragonal to cubic phase at between 42 and 57 °C.[10] However, the structural change between the two phases are not significant with only slight titling of octahedral framework.[11] While MAPbI$_3$-based perovskite solar cells have achieved high

PCEs, these cells are not stable in ambient conditions.[12] Their poor thermal stability also limits the cell module processing temperature.[13] At high temperatures, methylammonium metal halide perovskites are easily dissociated to organic halides and lead halides, degrading the photovoltaic performance.[14]

Replacing the organic cation with cesium (Cs) is considered as one possible way to address the aforementioned issues.[15-16] The solar cells based on all-inorganic cesium lead iodide ($CsPbI_3$) perovskite quantum dots (QDs) have been shown to achieve the highest certified PCE of 13.43% among QD solar cells,[17] and optically-pumped vertical cavity lasers incorporating $CsPbBr_3$ QDs have been shown to achieve ultra-low lasing threshold.[18] Both works reported outstanding stability among perovskite optoelectronic devices.

Thus far, most of perovskite thin film fabrications utilize solution-processing methods such as spin-coating and drop-casting. However, $CsPbI_xBr_{3-x}$ is much less soluble in solutions, which limits its applications in solution-processed optoelectronic devices.[19] On the other hand, the vapor-deposition method offers many advantages including good film morphology, precisely controlled thickness, ease of patterning, high reproducibility and better adaptability in flexible device fabrications.[20-21] In this work, we reported for the first time a systematic study on the structural phase transitions of vapor-deposited all-inorganic perovskite $CsPbIBr_2$ thin films. The as-deposited inorganic perovskite exhibits the perovskite (PVSK) phase and can be transitioned to the non-perovskite (non-PVSK) phase after appropriate water moisture exposure. This nonvolatile phase transition can be reversed repeatedly through heat/moisture exposure treatment cycles. The two different phases correspond to significant distinctions in the material's optoelectronic properties including photoluminescence (PL), optical absorption, charge transport. To demonstrate potential applications utilizing such characteristics, we also fabricated a photodetector (PD) array by vapor-depositing the CsPbIBr$_2$ film on a substrate pre-patterned with inkjet-printed Ag interdigitated electrodes (IDEs). In addition to heating all PD pixels uniformly on a heater platform, we can use a Nd:YVO$_4$ laser with 1064 nm wavelength to generate heat selectively in specific PD pixels through the photothermal effect in Ag,[22] which makes the transition from the non-PVSK to PVSK phase site-selectable. The PD pixel in the PVSK phase shows high responsivity up to 1.5 A/W with a broad absorption spectrum from 350 to 600 nm. On the contrary, the PD pixel in the non-PVSK phase exhibits a much smaller photoresponsivity and a narrower absorption spectrum. The large difference in the PD performance between these two nonvolatile phases provides the opportunities for potential applications in optical memory and information storage. The laser heating and moisture exposure can be used to "set" and "reset" the selected PD pixels, and the photocurrent is read out to extract the phase information patterned by laser writing. In addition, the PD array in the PVSK phase presents a possibility for flexible image sensors, and imaging of a pre-designed pattern is demonstrated using a flexible perovskite PD array.

**Results and discussion**

**1. Characterization of vapor-deposited CsPbIBr$_2$ thin films**

In this study, we used CsPbIBr$_2$ in further characterizations and device fabrication due to its better stability. The fabrication procedure of vapor-deposited CsPbIBr$_2$ PD array on rigid glass or flexible PET substrates is schematically illustrated in **Figure 1**a. The substrate was first ultrasonically cleaned and placed onto a printing platform. Conductive silver ink was then inkjet-printed onto the substrate to form an 8×8 interdigitated electrode (IDE) array (see METHODS section). The CsPbIBr$_2$ film was deposited onto the prepatterned substrate via a vapor-deposition process. The CsI, PbBr$_2$, were separately placed into 2 crucibles and simultaneous sublimed. By controlling the evaporation rates of the two sources, the mixed halide inorganic CsPbIBr$_2$ films can be obtained. Energy dispersive (EDS) X-ray analysis was used to estimate the halide stoichiometry of the resulting inorganic perovskite films. Figure S1 (see Supporting Information) shows the element distribution of iodide (I) and bromide (Br), the I/Br molar ratio maintains approximately 1:2 along the feature line, indicating formation of the CsPbIBr$_2$ thin film.

The top-view scanning electron microscopy (SEM) image of the CsPbIBr$_2$ film is presented in Figure 1b. The homogenous CsPbIBr$_2$ film shows complete surface coverage without obvious pinholes. The atomic force microscopy (AFM) image of the CsPbIBr$_2$ film in Figure 1c shows the surface roughness of 4.1 nm in a typical area of 5×5 μm$^2$, smaller than that of spin coated film (see Figure S3), further confirming the flat and smooth surface morphology. A unique advantage of the vapor-deposition over solution-cast process is that patterned perovskite thin films can be readily achieved through a shadow mask and more compatible with flexible substrates. A dot array pattern made of vapor-deposited CsPbIBr$_2$ on a flexible PET substrate is clearly observed in Figure 1d.

## 2. Phase transitions of inorganic perovskites via heating and moisture exposure

As shown in **Figure** 2a, the as-deposited CsPbIBr$_2$ film (PVSK phase) displayed an orange-red color. The synthesized CsPbIBr$_2$ film could maintain its color in an inert or low humidity (<50%) environment when quickly cooled down to room temperature. Lin et.al. have reported the reversible phase transition of solution-cast $CsPbI_xBr_{3-x}$ films between the room-temperature non-perovskite phase and high-temperature perovskite phase.[11] A similar phenomenon was observed in the vapor-deposited $CsPbIBr_2$ film. After exposed to an atmosphere with 80% relative humidity (RH) for 30 minutes, the orange-red $CsPbIBr_2$ film gradually turned to be transparent. Reheating the transparent $CsPbIBr_2$ film (non-PVSK phase) around 160 ºC in a $N_2$-filled glovebox triggered the color-change from transparent back to orange-red again. The reversible phase transition of $CsPbIBr_2$ is repeatable and the optical properties in both phases remained consistent after 20 cycles, as discussed below.

Figure 2b presents the absorption spectra of the non-PVSK and PVSK phase $CsPbIBr_2$ films, showing the latter have significantly stronger optical absorption. Furthermore, the PVSK phase film shows an absorption edge at around 600 nm, while this is blue-shifted to 430 nm for the non-PVSK phase film. Such a distinction can be used to distinguish $CsPbIBr_2$ films in these two different phases. The reversibility of structural transitions between non-PVSK and PVSK phases was monitored by measuring the absorption spectra after repetitive moisture exposure and heating. Figure 2c shows the absorption of $CsPbIBr_2$ films in the two phases at 500 nm. The data exhibits good consistency over 20 repetitive moisture/heat treatment cycles.

The X-ray diffraction (XRD) technique was used to analyze the crystal structural change in more details (Figure 2d). We notice that the PVSK phase $CsPbIBr_2$ film exhibits characteristic peaks located at 14.86º, 21.24º, 30.0º, corresponding to the (100), (110), (200) planes of the $CsPbIBr_2$ cubic structure. This confirms that $PbBr_2$ and $CsI$ layers have reacted to form one cubic-phase perovskite layer through an inter-diffusion process. We also observe a small peak (denoted by *) located at 11.4°, indicative of the remaining PbBr$_2$, which is similar to the finding from Hutter et.al.[9] The XRD pattern of the moisture-exposed CsPbIBr$_2$ film displays main peaks at 10.3º, 13.7º, 18.6º, 23.6º, 27.6º, 28.8º, corresponding to (002), (012), (021), (023), (015), (032) planes of the CsPbIBr$_2$ orthorhombic phase.[23] Furthermore, we compared the XRD patterns of CsPbIBr$_2$ films in the two phases before and after phase-transition cycles. No obvious difference was observed for both phases after 20 cycles (Figure S4).

The different crystal structures of CsPbIBr$_2$ films in non-PVSK and PVSK phases also results in noticeable refractive index difference. We measured the complex refractive index using an ellipsometry, Figure 2e shows the real part (n) and imaginary part (k) of the complex refractive index of non-PVSK and PVSK phase CsPbIBr$_2$ films. The k value of the PVSK phase film is almost zero after 600 nm, indicating no optical loss for λ >600 nm. In contrast, the k value of the non-PVSK phase film cuts off at ~430 nm. These results are consistent with the bandgap and absorption spectra in Figure 2b. The large difference of refractive index between two phases in the visible light region may have potential applications in optical modulators, switches, tunable metasurfaces and other photonic devices.[24-25] Figure 2f presents the current-voltage (I-V) characteristics of a hole-only device (ITO/PEDOT:PSS/CsPbIBr$_2$/Au), the CsPbIBr$_2$ layer is either in the non-PVSK or PVSK phase. Both *I-V* curve ($I \sim V^n$) shows the ohmic (n = 1), trap filling (n > 2) and child regions (n = 2). The trap filled voltage ($V_{TFL}$) of the non-PVSK phase film was higher than that of the PVSK phase film, indicating a higher trap density ($n_{trap}$).[26-27] This may be attributed to the release of halide atoms and increase of halide vacancies caused by the interaction of water molecules with Pb atoms.[28-29]

**3. The patterned non-PVSK to PVSK phase transition through laser direct writing**

Considering current technological trends toward flexible and wearable devices, a robust processing method such as heating for large-area flexible substrates is becoming more and more important.[30] Laser heating is a promising candidate for such an application. Compared to other more common thermal heating methods, laser heating has the advantage of area-selective rapid heating and cooling, better compatibility with flexible substrates, and processing scalability.[31-33] In this work, we employed a near infrared (1064 nm, Nd:YVO$_4$) laser for laser direct writing (LDW) of the non-PVSK phase CsPbIBr$_2$ film deposited on an interfacial layer (ITO or Ag). This method can precisely control the heating temperature and crystallization process without damage by the laser beam.[34] **Figure 3**a presents the schematic configuration of the LDW experimental setup. The sample was placed on a XY translation stage and driven laterally, a 1064 nm laser beam was focused onto the sample and laterally scanned the non-PVSK phase CsPbIBr$_2$ film on ITO. Two orange-red stripes were formed as shown in the inset photo of figure 3a, indicating an instantaneous transformation of non-PVSK to PVSK phase triggered by the laser beam.

Figure 3b displays the optical (top) and fluorescent (bottom) microscopy images of a LDW pattern on the perovskite film. The laser beam was focused into a ~ 50 μm spot and scanned to form patterns. The PL of the PVSK phase area is much stronger than that of the non-PVSK phase area, which is clearly manifested in these fluorescent images. Figure 3c presents the line-scan PL intensity of the micro-line array pattern in Figure 3b, demonstrating the uniform periodic fluorescence emitted from the PVSK phase micro-line array. Raman spectroscopy has high positional and spectral accuracy, it was used here to investigate the phase mapping of the LDW patterns. In Figure 3d, the top Raman spectrum (blue curve) corresponds to the micro-lines with strong red fluorescence. A prominent peak at 138 cm$^{-1}$ is observed, which is attributed to the PVSK phase.[35] On the other hand, Raman spectrum of the dark background area (bottom black curve) shows a red-shifted and broader peak at 114 cm$^{-1}$, which corresponds to the non-PVSK phase.[32] The Raman measurement finding matches well with other reported literature,[36-38] further demonstrating the phase transformation induced by LDW.

The COMSOL simulated temperature distribution around a NIR laser spot (1 um diameter) shown in Figure 3e clarifies the underlying mechanism of photothermal heating. The NIR laser beam is not absorbed by the perovskite layer due to its wavelength longer than the absorption edge, and the photothermal effect from the interfacial layer (ITO) contributes to rapid local heating. A relative low laser power (9 mW) could lead a temperature rise of 160 ºC localized to the focused beam position, which is hot enough to locally induce the phase transition. Around the focused laser spot, the temperature decreased to half of the maximum at a distance 1.4 μm from the center (see Figure S6 for thermal simulation in details). The increased surface temperature as a function of laser intensity was measured using an IR thermal camera for both Ag and ITO as the interfacial layer (see Figure S7a). The temperature rise strongly depends on the material of the interfacial layer, and it was observed that the Ag layer induced more photothermal heating and temperature rise than

ITO at the same laser intensity. To induce phase transition from the non-PVSK to PVSK phase (160 °C), the laser intensity for Ag and ITO is ~115 and ~235 W/cm$^2$ respectively.

**4. Optical memory device for data storage enabled by non-volatile phase transitions**

The as-fabricated PVSK phase PD array was first exposed to water moisture with relative humidity of 80% until the perovskite film became transparent. As schematically illustrated in **Figure 4**a, one PD pixel was then heated by the focused laser beam selectively. The induced photothermal heating led to an instantaneous temperature rise at specific pixels, and a color change from transparent to orange-red was observed within a second. The inset photo in Figure 4a is a fluorescent image of one pixel heated by the laser beam. The uniform fluorescence inside the IDE demonstrate the occurrence of phase transformation in that pixel. We noticed that the fluorescence is weaker directly on top of the Ag electrode area, which is attributed to PL quenching effect by Ag.

We then measured the current-voltage (*I-V*) curves of both non-PVSK and PVSK phase PD pixels in dark condition and under white light illumination (tungsten-halogen lamp), the results are presented in Figure 4b. PDs in two phases exhibit similar dark current ($I_{dark}$) on the level of 10$^{-9}$ A at 5 V. The non-PVSK phase PD has slightly higher dark current, which may be attributed to the higher conductivity of non-PVSK phase films as shown in Figure 2f. Upon white light illumination, the photoinduced carriers contributed to the current increase of PDs in both phases. However, the light current ($I_{light}$) of PDs in PVSK phase was ~50 times larger than that of the PDs in non-PVSK phase when operated at a bias of 5 V. The large difference of photocurrents between PDs in two phases is attributed to the different crystal structures and optical absorptions. The responsivity ($R$), a key figure of merit for a PD, is given by[39]

$$R = \frac{I_{ph}}{E_e S} = \frac{I_{light} - I_{dark}}{E_e S}$$

where $E_e$ and $S$ are the illumination intensity and the single PD device area, respectively. As shown in Figure 4c, a large difference of spectral responsivity between PDs in the two phases is observed. The responsivity spectrum shows a cut-off at 430 nm and 600 nm for the non-PVSK and PVSK phase PDs, respectively. Since the PVSK phase PDs have much better photodetection performance in terms of responsivity and photodetection wavelength range compared to the non-PVSK phase PDs, we then focused on systematically investigating the PVSK phase PD performance.

The *I-V* curves of the PVSK phase PDs under 405 nm irradiation with light intensity varying from 0 to 50 mW/cm² are shown in Figure 4d, the current shows linear relationship with voltage. Response speed is another key parameter for photodetectors. Figure 4e exhibits the time-dependent photocurrent response under periodically modulated UV light illumination. The PDs show stable and reproducible photo-response behaviors under periodic on/off switching cycles. The rise time (defined as the time for the photocurrent to increase from 10% to 90% of the peak value) and the fall time (from 90% to 10% of the peak value) are extracted to be 0.8 and 1 ms (inset of Figure 4f), respectively. We then measured the 3dB frequency bandwidth by recording photocurrent as a function of the modulated light frequency, as presented in Figure 4f. The 3dB bandwidth $f_{3dB}$ is found to be ~400 Hz, which is consistent with theoretical value calculated by $f_{3dB} = 0.35 / t_{rise}$.

As discussed above, the crystal structure difference between the two phases leads to dramatically different photo-response under white light irradiation, which can be used to distinguish these two phases. Considering both non-PVSK and PVSK phases are nonvolatile in moderate humidity (RH<50%) environment, the CsPbIBr$_2$ PD array can be potentially applied as a rewritable optical memory device. **Figure 5**a schematically illustrates the working mechanism of this application. The laser beam directly writes the individual PD pixels to transform the non-PVSK phase to PVSK phase, which represents the 'write' or 'set' process. The moisture exposure triggers the phase transition from the PVSK to non-PVSK phase again, which is the 'erase' or 'reset' process. Mapping of the pixels that have been written (set) can be obtained through reading the photocurrent of each PD pixel under a large-area white light irradiation (the 'read' process). The pixel with a high photocurrent stores '1' and the pixel with a low photocurrent stores '0', then the optical information encoded by the laser beam writing can be restored. Figure S7b (see Supporting Information) shows a PD array photo where the first-row pixels (8 pixels) were heated by NIR laser beam and displayed orange-red color.

We also characterized the set/reset cycle endurance and the data retention ability of the PD pixels as optical memories. The photocurrents of the perovskite PD pixels over repetitive laser heating/moisture exposure cycles are presented in Figure 5b. The ratio of the photocurrent from the PD pixel in the PVSK phase to that in the non-PVSK phase was observed to maintain above 50 over 10 cycles. This is consistent with the reversible absorption properties of the perovskite layer shown in Figure 2c and is high enough to distinguish the '1' state from the '0' state. The data retention capability was verified by recording the photocurrent continuously over time. As shown in Figure 5c, the photocurrents of the PD pixels in both phases remain relatively stable. About 20% degradation for the PVSK phase PD was observed after being stored for 16 hours in our lab (RH 45%). The stability can be further improved by reducing the environment humidity during test. These findings show that the perovskite PD array can be used to conduct write-read-erase cycles many times, suggesting the potential of utilizing the non-volatile phase-change properties of vapor-deposited $CsPbIBr_2$ for optical memory and information storage applications.

**5. Imaging application of the flexible perovskite PD arrays**

Through inkjet printing and vapor deposition technologies, high-quality perovskite PD arrays can be fabricated on large-area and flexible substrates without requiring extra-high annealing temperature. These advantages can enable scalable device fabrication and facilitate commercialization of flexible perovskite PD arrays for image sensor applications. Herein, we use a 16×16 PD pixel array to verify the imaging capability as proof-of-concept demonstration. The optical image of the 16×16 PD pixel array is presented in Figure S9b. The as-deposited flexible PD array on PET through the vapor-deposition process was tested without further annealing. In an array of PDs, it is important to examine the uniformity of the photo-response from all 16×16 pixels. A uniform white light source was projected onto the PD array, and the photo-response from all 256 pixels was measured and mapped. **Figure 6**a presents the photocurrent distribution along selected four lines. The photocurrent fits in a tight range of 0.14 to 0.15 µA, with only 2% fluctuation relative to the average value, demonstrating excellent uniformity over all PD pixels.

The imaging capability of the flexible perovskite PD array was verified by reconstructing a projected optical pattern. Figure 6b shows the experimental setup composed of a collimated white light source, a photomask with the desired pattern and the flexible perovskite PD array device. After scan-measuring the light current of each pixel, we obtain the light current distribution of all 256 pixels. Each pixel is given a gray scale number ($G$) between 0 and 1 based on the measured light current ($I_{meas}$) through the following equation.[40]

$$G = \frac{I_{meas} - I_{dark}}{I_{max} - I_{dark}}$$

Where $I_{dark}$ is the average dark current and $I_{max}$ is the maximum light current of all 256 pixels. As shown in Figure 6c, a "UW" pattern was projected onto the PD array and retrieved from the photocurrent mapping. The performance of the flexible perovskite PD array was also characterized at different bending angles and after various numbers of bending cycles. The results show little degradation after bending (Figure S8), supporting the promise of vapor-deposited perovskite films for flexible devices.

Since the response speed ($t_{rise}$= 0.8 ms) of our vapor-deposited perovskite PD arrays is faster than the human-eye recognition speed (>40 ms),[20] our demonstrations open up the possibility of using vapor-deposited inorganic perovskite thin film and inkjet printing technology for flexible, large-scale, high-speed image acquisition in a broad spectrum of applications such as wearable electronic eyes and cameras.

**Conclusions**

In summary, we have demonstrated uniform inorganic perovskite films with excellent optoelectronic properties grown by a vapor-deposition process. The fabrication method can be easily implemented with various substrates. The reversible phase transition between non-PVSK and PVSK phases was observed and confirmed by XRD. The two phases manifest significant differences in optical absorption, refractive index and other optoelectronic properties. Utilizing the phase-change properties, the fabricated perovskite PD array can be applied as a rewritable optical memory device for information storage. Finally, we demonstrated optical imaging using flexible perovskite PD array devices on PET substrates. These works suggest the versatile utilities of vapor-deposited inorganic perovskite films through the unique reversible phase-changing features and the potential for fabrication of flexible, large-area devices.

**Experimental Section**

*Materials*: CsI (99.9% trace metal purity), $PbBr_2$ (99.99% trace metal purity) and the Ag nanoparticle ink (Silverjet DGP-40LT-15C) for inkjet printing were purchased from Sigma-Aldrich and used without further purification. The evaporation crucibles were purchased from Kurt J. Lesker.

*Inkjet printing Ag electrodes*: PET and glass substrates were first cleaned by ultrasonication with 2-propanol and DI water in sequence. The Ag nanoparticle ink was loaded into the cartridge of an inkjet printer (Diamtrix), the cartridge was ultrasonicated briefly to reduce the aggregation of Ag nanoparticles. The voltage of printing nozzle was set at 23 V, the waveform frequency was set at 5 kHz. The cleaned substrates were well adhered to the printing platform using vacuum, the platform temperature was set at room temperature. The printing nozzle was controlled by the computer with the designed pattern to drop the Ag ink with a proper drop spacing distance (30 μm for PET and 40 μm for glass). The substrates were then heated in oven at 110 °C for 10 minutes to sinter the Ag electrodes.

*Perovskite vapor deposition*: Inorganic perovskite $CsPbI_xBr_{3-x}$ was deposited by dual source thermal deposition method. For $CsPbIBr_2$ deposition, CsI and $PbBr_2$ powder are placed into two crucibles separately. The vacuum chamber was pumped down to $10^{-6}$ Torr. The evaporation rates of CsI and $PbBr_2$ were 0.53 and 0.5 A/s based on the calculation from their molecular mass and density. Two crucibles were heated simultaneous to appropriate temperature and the evaporation rates were monitored by two quartz crystal microbalances (QCMs) separately, the actual evaporation rates during the deposition process were controlled within 20% away from set values. Two QCMs were calibrated separately before deposition to determine the tooling factors for CsI and $PbBr_2$.

*Laser direct writing process*: Laser heating was performed using our homemade system based on a $Nd:YVO_4$ 1064 nm laser source (Spectra Physics). The sample was placed on a XY translation stage, the laser beam was focused through a 10x objective lens at the specific location of the sample. The photothermal effect on ITO or Ag layer induced by laser beam could increase the localized temperature and achieve the phase transition.

*Reversible phase transition of $CsPbIBr_2$*: The non-PVSK phase $CsPbIBr_2$ film was heated on hotplate (160 °C) for 10 minutes or by the 1064 nm laser beam to transform into the PVSK phase $CsPbIBr_2$ film. To convert the PVSK phase to non-PVSK phase, the $CsPbIBr_2$ film was placed in a humidity control box with relative humidity of 80 % for 30 minutes. The reversible phase conversion process was carried out 20 times.

*Perovskite thin film characterization*: The absorption spectra of perovskite films were acquired using a Varian Cary 5000 UV-vis-NIR Spectrophotometer. XRD patterns were obtained using Bruker D8 with a Cu Kα radiation ($\lambda$ = 1.54184 Å). The surface morphology of perovskite films was measured by SEM (FEI Sirion) and AFM (Bruker Icon). The refractive index was measured using an ellipsometer (J.A. Woollam M2000), the raw data was fitted by the B-spline model on a Cauchy dispersion glass substrate using Complete EASE software. Raman spectra were acquired by a Raman microscope (InVia, Renishaw), 785 nm laser was used as the excitation source to avoid the interruption of fluorescence from sample. The perovskite was deposited on the inkjet-printed Ag layer to enhance the Raman scattering intensity. The fluorescent images were obtained through a fluorescence microscope (EVOS, Thermo Fisher Scientific), the raw images were processed by ImageJ software to adjust the threshold intensity and remove background signal.

*Perovskite PD array measurement*: Current-voltage characteristics of perovskite photodetectors both in dark and under light illumination were measured by a source meter (Keithley 6430). The irradiation light was provided by a 405 nm laser source, the light intensity could be easily adjusted. The time-dependent current response was recorded under periodically on-off light illumination from a 405 nm laser source driven by the square wave from a signal generator. The spectral EQE measurement was performed by using a tungsten-halogen light source to illuminate the device through a monochromator (Acton Research SpectraPro 275), which provides the monochromatic light excitation. The frequency response of the photodetector was measured by modulating the 405 nm laser light at different frequencies using an optical chopper. The generated photocurrent was recorded using a lock-in amplifier (Stanford Research SRS 830).

For image sensor application, the perovskite PD array was placed underneath an aluminum mask with pre-designed pattern. The collimated white light from a tungsten-halogen lamp went through the pattern area of the mask and arrives at the PD array platform. The currents of 16×16 PD pixels were recorded and mapped.

**Author Information**

Corresponding author
*Email: lylin@uw.edu

ORCID
Chen Zou: 0000-0001-9638-6363
Lih Y. Lin: 0000-0001-9748-5478

Notes
The authors declare no competing financial interest.

**Supporting Information**
Supporting Information is available from the Wiley Online Library or from the author.

**Acknowledgement**
This work is supported in part by National Science Foundation (grant ECCS-1807397 and CHE-1836500). Part of the work was conducted at the Molecular Analysis Facility, a National Nanotechnology Coordinated Infrastructure site at the University of Washington which is supported in part by the National Science Foundation (grant NNCI-1542101), the University of Washington, the Molecular Engineering & Sciences Institute, the Clean Energy Institute, and the National Institutes of Health. C. Z. would like to thank Felippe J. Pavinatto from the UW Clean Energy Institute Testbed (WCET) for discussion and help with inkjet printing.

References

[1]	B. R. Sutherland, E. H. Sargent, *Nat. Photonics* **2016**, *10*, 295.
[2]	T. M. Brenner, D. A. Egger, L. Kronik, G. Hodes, D. Cahen, *Nat. Rev. Mater.* **2016**, *1*.
[3]	A. Kojima, K. Teshima, Y. Shirai, T. Miyasaka, *J. Am. Chem. Soc.* **2009**, *131*, 6050.
[4]	C. S. Ponseca, T. J. Savenije, M. Abdellah, K. Zheng, A. Yartsev, T. Pascher, T. Harlang, P. Chabera, T. Pullerits, A. Stepanov, J.-P. Wolf, V. Sundström, *J. Am. Chem. Soc.* **2014**, *136*, 5189.
[5]	M. A. Green, A. Ho-Baillie, H. J. Snaith, *Nat. Photonics* **2014**, *8*, 506.
[6]	W. S. Yang, B. W. Park, E. H. Jung, N. J. Jeon, Y. C. Kim, D. U. Lee, S. S. Shin, J. Seo, E. K. Kim, J. H. Noh, S. I. Seok, *Science* **2017**, *356*, 1376.
[7]	S. D. Stranks, H. J. Snaith, *Nat. Nanotechnol.* **2015**, *10*, 391.
[8]	M. E. Ziffer, D. S. Ginger, *Science* **2016**, *353*, 1365.
[9]	E. M. Hutter, R. J. Sutton, S. Chandrashekar, M. Abdi-Jalebi, S. D. Stranks, H. J. Snaith, T. J. Sayenije, *ACS Energy Lett.* **2017**, *2*, 1901.
[10]	R. L. Milot, G. E. Eperon, H. J. Snaith, M. B. Johnston, L. M. Herz, *Adv. Funct. Mater.* **2015**, *25*, 6218.
[11]	J. Lin, M. L. Lai, L. T. Dou, C. S. Kley, H. Chen, F. Peng, J. L. Sun, D. L. Lu, S. A. Hawks, C. L. Xie, F. Cui, A. P. Alivisatos, D. T. Limmer, P. D. Yang, *Nat. Mater.* **2018**, *17*, 261.
[12]	Y. Rong, L. Liu, A. Mei, X. Li, H. Han, *Adv. Energy Mater.* **2015**, *5*, 1501066.
[13]	R. J. Sutton, G. E. Eperon, L. Miranda, E. S. Parrott, B. A. Kamino, J. B. Patel, M. T. Horantner, M. B. Johnston, A. A. Haghighirad, D. T. Moore, H. J. Snaith, *Adv. Energy Mater.* **2016**, *6*, 1502458.
[14]	G. E. Eperon, G. M. Paterno, R. J. Sutton, A. Zampetti, A. A. Haghighirad, F. Cacialli, H. J. Snaith, *J. Mater. Chem. A* **2015**, *3*, 19688.
[15]	M. Kulbak, D. Cahen, G. Hodes, *J. Phys. Chem. Lett.* **2015**, *6*, 2452.
[16]	J. Liang, C. Wang, Y. Wang, Z. Xu, Z. Lu, Y. Ma, H. Zhu, Y. Hu, C. Xiao, X. Yi, *J. Am. Chem. Soc.* **2016**, *138*, 15829.
[17]	A. Swarnkar, A. R. Marshall, E. M. Sanehira, B. D. Chernomordik, D. T. Moore, J. A. Christians, T. Chakrabarti, J. M. Luther, *Science* **2016**, *354*, 92.
[18]	C. Y. Huang, C. Zou, C. Y. Mao, K. L. Corp, Y. C. Yao, Y. J. Lee, C. W. Schlenker, A. K. Y. Jen, L. Y. Lin, *ACS Photonics* **2017**, *4*, 2281.
[19]	C. Y. Chen, H. Y. Lin, K. M. Chiang, W. L. Tsai, Y. C. Huang, C. S. Tsao, H. W. Lin, *Adv. Mater.* **2017**, *29*.
[20]	G. Tong, H. Li, D. Li, Z. Zhu, E. Xu, G. Li, L. Yu, J. Xu, Y. Jiang, *Small* **2018**, *14*, 1702523.
[21]	Y. Hu, Q. Wang, Y.-L. Shi, M. Li, L. Zhang, Z.-K. Wang, L.-S. Liao, *J. Mater. Chem. C* **2017**, *5*, 8144.
[22]	W. Fang, J. Yang, J. Gong, N. Zheng, *Adv. Funct. Mater.* **2012**, *22*, 842.
[23]	M. Lai, Q. Kong, C. G. Bischak, Y. Yu, L. Dou, S. W. Eaton, N. S. Ginsberg, P. Yang, *Nano Research* **2017**, *10*, 1107.
[24]	J. Zheng, A. Khanolkar, P. Xu, S. Colburn, S. Deshmukh, J. Myers, J. Frantz, E. Pop, J. Hendrickson, J. Doylend, N. Boechler, A. Majumdar, *Opt Mater Express* **2018**, *8*, 1551.

[25]    C.-H. Liu, J. Zheng, S. Colburn, T. Fryett, Y. Chen, X. Xu, A. Majumdar, *Nano Lett.* **2018**, *18*, 6961.
[26]    Q. Dong, Y. Fang, Y. Shao, P. Mulligan, J. Qiu, L. Cao, J. Huang, *Science* **2015**, *347*, 967.
[27]    M. I. Saidaminov, A. L. Abdelhady, B. Murali, E. Alarousu, V. M. Burlakov, W. Peng, I. Dursun, L. F. Wang, Y. He, G. Maculan, A. Goriely, T. Wu, O. F. Mohammed, O. M. Bakr, *Nat. Commun.* **2015**, *6*, 7586.
[28]    E. Mosconi, J. M. Azpiroz, F. De Angelis, *Chem. Mater.* **2015**, *27*, 4885.
[29]    A. Mattoni, A. Filippetti, C. Caddeo, *J. Phys. Condens. Matter* **2016**, *29*, 043001.
[30]    S. S. Chou, B. S. Swartzentruber, M. T. Janish, K. C. Meyer, L. B. Biedermann, S. Okur, D. B. Burckel, C. B. Carter, B. Kaehr, *J. Phys. Chem. Lett.* **2016**, *7*, 3736.
[31]    S. J. Kim, J. Byun, T. Jeon, H. M. Jin, H. R. Hong, S. O. Kim, *ACS Appl. Mater. Interfaces* **2018**, *10*, 2490.
[32]    J. A. Steele, H. F. Yuan, C. Y. X. Tan, M. Keshavarz, C. Steuwe, M. B. J. Roeffaers, J. Hofkens, *ACS Nano* **2017**, *11*, 8072.
[33]    I. Konidakis, T. Maksudov, E. Serpetzoglou, G. Kakavelakis, E. Kymakis, E. Stratakis, *ACS Appl. Energy Mater.* **2018**, *1*, 5101.
[34]    T. Jeon, H. M. Jin, S. H. Lee, J. M. Lee, H. I. Park, M. K. Kim, K. J. Lee, B. Shin, S. O. Kim, *ACS Nano* **2016**, *10*, 7907.
[35]    S. M. Jain, B. Philippe, E. M. Johansson, B.-w. Park, H. Rensmo, T. Edvinsson, G. Boschloo, *J. Mater. Chem. A* **2016**, *4*, 2630.
[36]    L. Q. Xie, T. Y. Zhang, L. Chen, N. Guo, Y. Wang, G. K. Liu, J. R. Wang, J. Z. Zhou, J. W. Yan, Y. X. Zhao, B. W. Mao, Z. Q. Tian, *Phys. Chem. Chem. Phys.* **2016**, *18*, 18112.
[37]    M. Ledinsky, P. Loper, B. Niesen, J. Holovsky, S. J. Moon, J. H. Yum, S. De Wolf, A. Fejfar, C. Ballif, *J. Phys. Chem. Lett.* **2015**, *6*, 401.
[38]    R. G. Niemann, A. G. Kontos, D. Palles, E. I. Kamitsos, A. Kaltzoglou, F. Brivio, P. Falaras, P. J. Cameron, *J. Phys. Chem. C* **2016**, *120*, 2509.
[39]    C. Zou, Y. Xi, C. Y. Huang, E. G. Keeler, T. Feng, S. Zhu, L. D. Pozzo, L. Y. Lin, *Adv. Opt. Mater.* **2018**, *6*, 1800324.
[40]    L. Gu, M. M. Tavakoli, D. Zhang, Q. Zhang, A. Waleed, Y. Xiao, K. H. Tsui, Y. Lin, L. Liao, J. Wang, *Adv. Mater.* **2016**, *28*, 9713.

**Figure Captions**

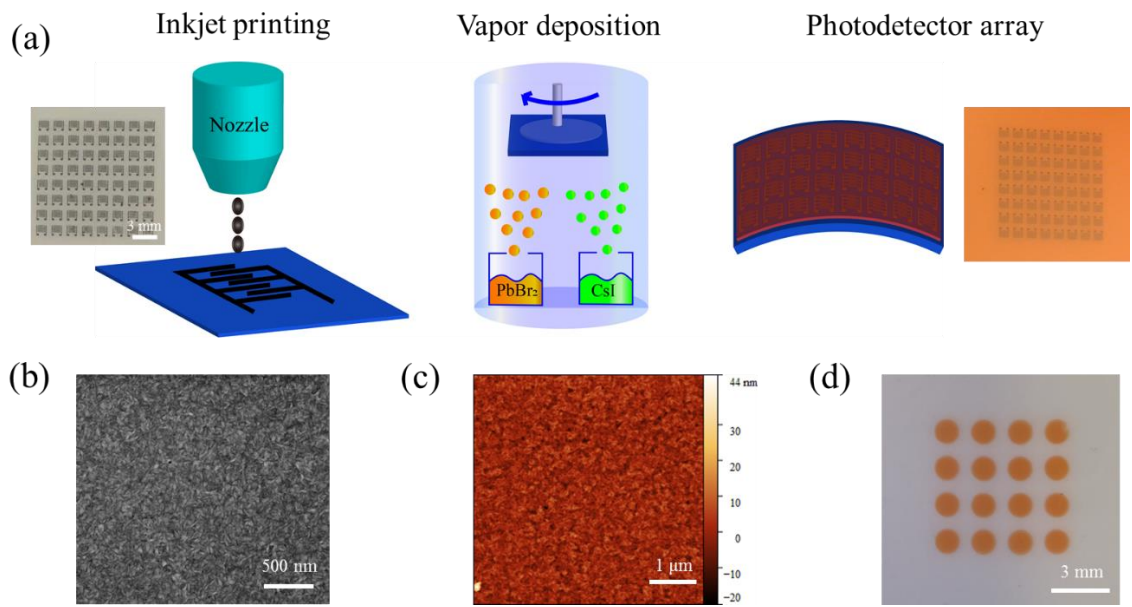

**Figure 1**. Fabrication and characterization of all-inorganic $CsPbIBr_2$ perovskite thin films. a) The schematic fabrication procedure of all-inorganic perovskite photodetector array via inkjet printing technology and dual-source thermal vapor-deposition method b) The SEM image of the $CsPbIBr_2$ film with a scale bar of 500 nm. c)The AFM image of the $CsPbIBr_2$ film, the RMS roughness is 4.1 nm. d) The optical image of a vapor-deposited perovskite dot array on a flexible PET substrate patterned through a shadow mask.

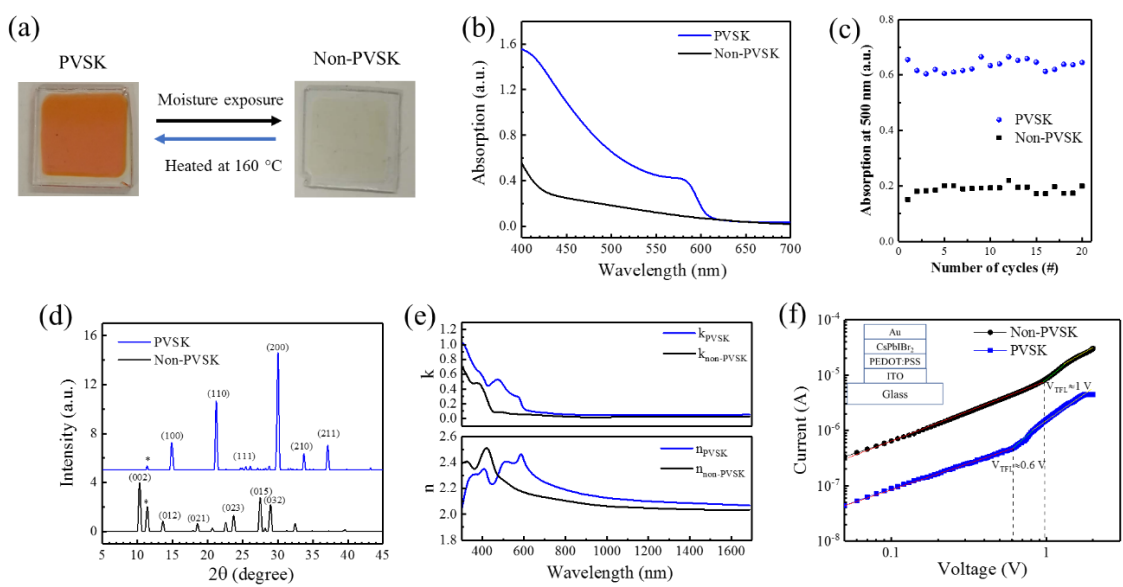

**Figure 2**. Phase transitions of inorganic perovskite thin films via laser heating and moisture exposure, taking CsPbIBr$_2$ for example. a) Optical images of PVSK (orange-red color) and non-PVSK phase (transparent) CsPbIBr$_2$ thin films. The reversible phase transitions can be achieved by heating/moisture exposure. b) The absorption spectra of PVSK and non-PVSK CsPbIBr$_2$ films. c) The reversible and repeatable switching behavior of the absorption (at 500 nm) of the CsPbIBr$_2$ thin film over 20 phase transition cycles. d) XRD patterns and e) complex refractive index spectra of PVSK and non-PVSK CsPbIBr$_2$ films. f) *I-V* curves of hole only devices (ITO/PEDOT:PSS/CsPbIBr$_2$/Au) where the CsPbIBr$_2$ layer is either in the non-PVSK or PVSK phase. The *I-V* curve is fitted by $I \sim V^n$, the red, green and yellow fitting line represents n = 1, n > 2 and n = 2 respectively.

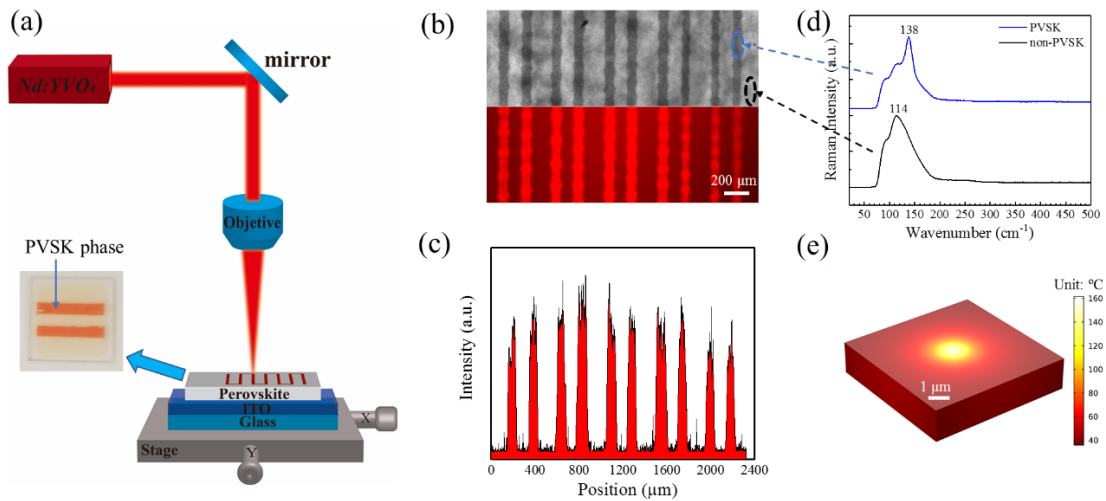

**Figure 3**. Laser direct writing (LDW) of the perovskite films assisted by photothermal effect. a) The schematic illustration for photothermal heating at the interfacial layer (ITO or Ag). The inset shows the photo of a perovskite thin film with two stripes converted to the PVSK phase through LDW. b) Optical (top) and fluorescent (bottom) microscopy images of the micro-lines pattern by LDW. c) The line-scan PL intensity of the fluorescent image in b). d) The measured Raman spectra at different positions. The two distinct spectra correspond to the non-PVSK phase (black curve) and the PVSK phase (blue curve). e) COMSOL Multiphysics thermal simulation of heat distribution induced by a laser beam (9 mW) with a spot size of 1 um.

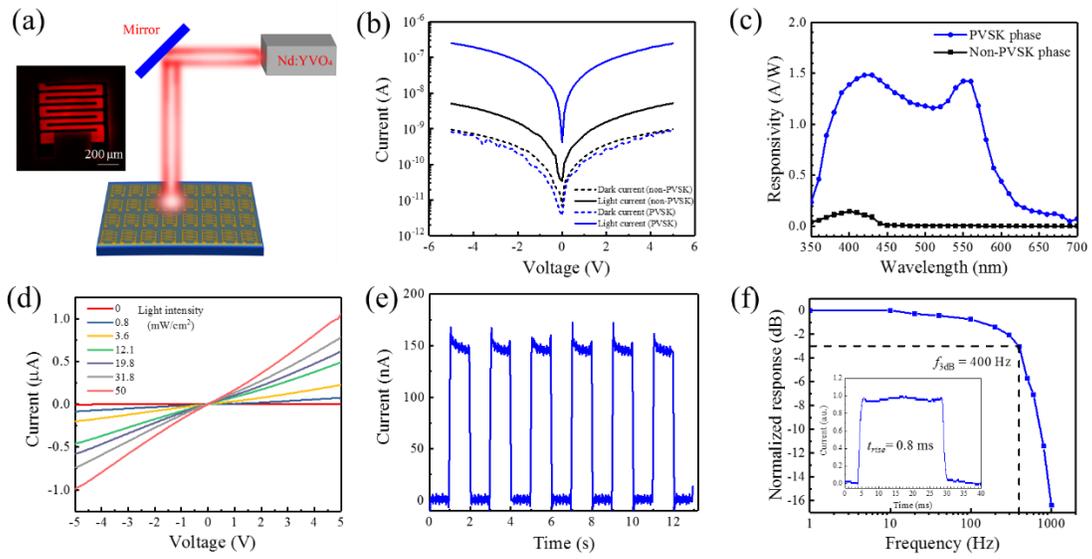

**Figure 4**. Detection performance of inorganic perovskite photodetector arrays. a) The schematic illustration of a single perovskite PD pixel heated by a 1064 nm laser. The inset photo is the fluorescent image of the pixel after transitioning to the PVSK phase. b) *I-V* curves of CsPbIBr$_2$ PD pixels in the non-PVSK phase and PVSK phase, under dark and white light illumination. c) The responsivity spectra of the perovskite PD pixel in two phases. d) *I-V* curves of the perovskite PD in the PVSK phase under 405 nm light illumination with different intensities. e) The time-dependent photocurrent switching behavior, and f) normalized frequency response of the photodetector in the PVSK phase. The inset shows the magnified photoresponse under 20 Hz light modulation.

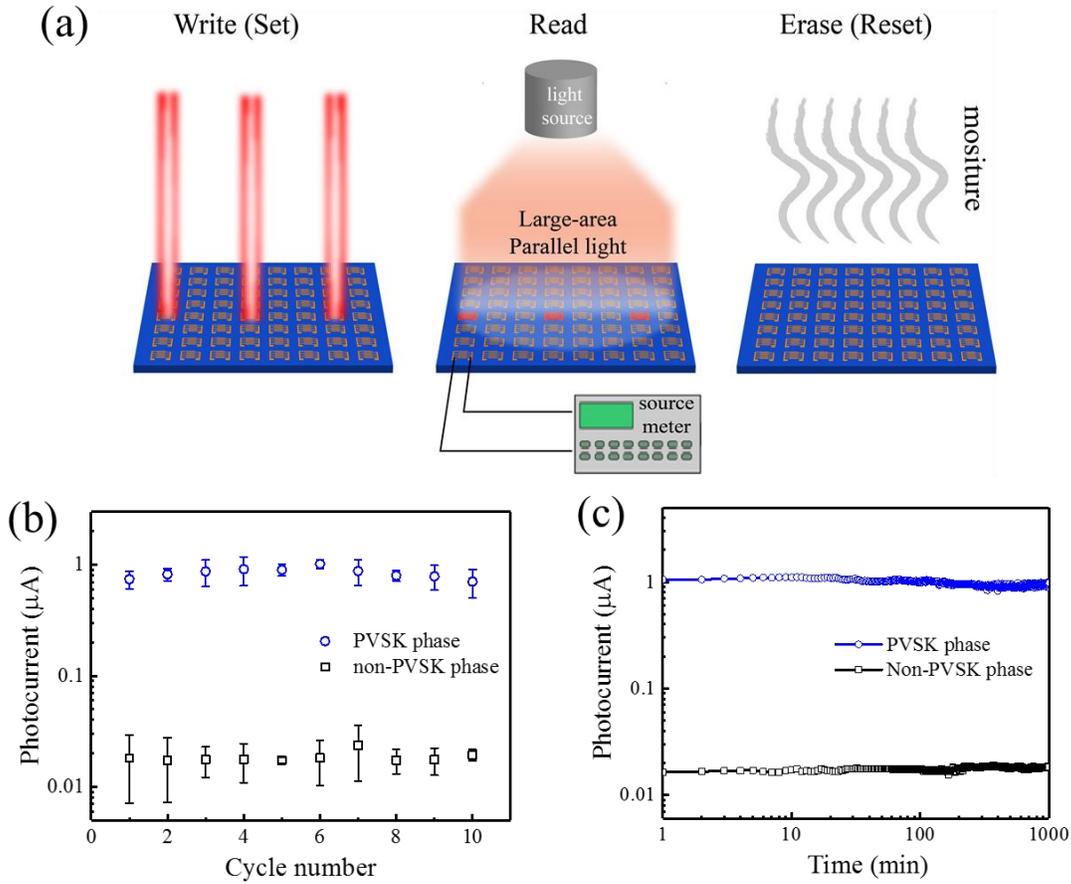

**Figure 5**. The optical memory application using the perovskite photodetector arrays enabled by the phase transition between the non-volatile PVSK and non-PVSK phases. a) Schematic illustration of the working principle. b) The photocurrents of perovskite PD pixels in PVSK and non-PVSK phases during 10 cycles of laser heating/moisture exposure test. c) Data retention capability of the non-volatile PVSK and non-PVSK phase perovskite PDs. The photocurrent only decreases to 80% of the original value for the PVSK phase after 1000 minutes under 45% relative humidity.

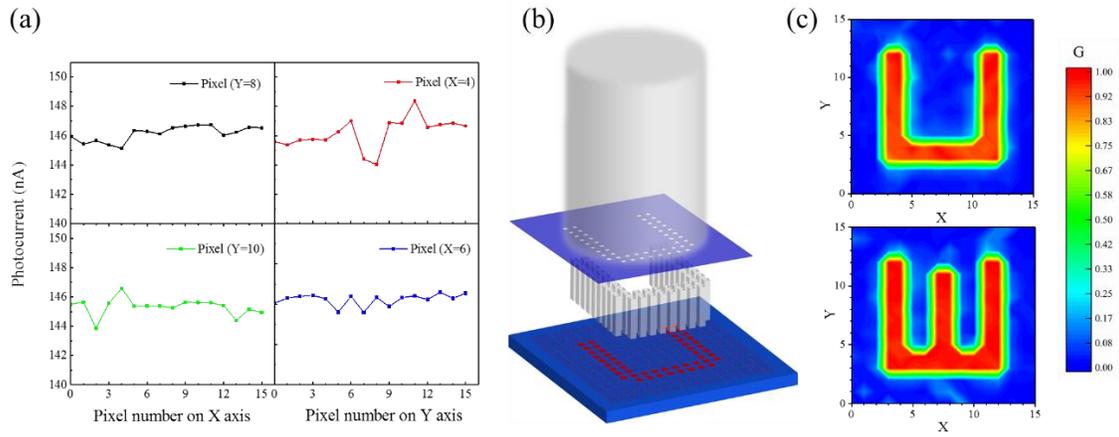

**Figure 6**. Imaging application of the flexible perovskite PD arrays. a) The line-scan photocurrent profiles measured from four randomly selected lines in a 16x16 array. b) The schematic setup of the imaging experiment. An optical pattern was projected onto the PD array through a prepatterned mask using a collimated white light source. c) a 'UW' pattern reconstructed by the PD array through mapping the photocurrent from each pixel.